# The Good, The Bad & The Ugly Features: A Meta-analysis on User Review About Food Journaling Apps

**Ahmed Fadhil**

University of Trento, Fondazione Bruno Kessler, Trento, Italy

*fadhil@fbk.eu*

**ABSTRACT**

Users review about an app is a crucial component for open mobile application market, such as the AppStore and the Google play. Analyzing these reviews can reveal user's sentiment towards a feature in the app. There exist several analytical tools to summarize user reviews and extract meaningful sense out of them. However, these tools are still limited in terms of expressiveness and accurately classifying the reviews into more than a positive and a negative review. There is a need to get more insights from user app reviews and direct it to future app development. In this paper, we present our result of analyzing user reviews of 20 food journaling and health tracking apps. We gathered and analyzed reviews per app and classified them into three distinct categories using the sentiment treebank with recursive neural tensor network. We then analyzed the vocabulary frequency per category using the Gensim implementation of Word2Vec model. The analysis result clustered the reviews into good, bad and ugly feature reviews. Different usage patterns were detected from users review. We identified major reasons why users express a certain sentiment towards an app and learned how users' satisfaction or complaints was related to a specific feature. This research could be a guideline for app developers to follow when developing an app to refrain from adopting techniques that might demotivate (hinder) the application use or adopt those perceived positively by the users.

## 1. INTRODUCTION

With the increasing burden of overweight and obesity epidemics and their associated health risk factors, several smartphone apps are developed to help maintain individuals' weight and promote their physical activity. These apps track various aspects of individuals health, including diet, physical activity and sleep. With healthcare reformation and data availability, mobile health apps are emerging as alternative resources over physicians [15]. Although the abundant apps installed, however it is impossible to keep up with all of them. According to a NYU Langone Medical Center1 study of 1,604 US smartphone owners in June 2015, 41%, downloaded more than five apps. Half of respondents, 41%, said they would not pay for a health app, while 20% said they would pay a maximum of $1.99, and 22 % would pay a maximum of between $2.00 and $5.99 [14]. A majority, 65%, of respondents that have downloaded at least one health app, opened it at least once a day, and 44% of this group used their app between 1 and 10 minutes. The study concluded that there is a huge cultural lack of understanding between developers and the consumers. Another study by IT-Online have shown 79% of Americans said they would use a wearable device to manage their health, whereas 45% wanted tracking of symptoms, and 43% wanted it to manage a personal health issue or condition [15]. As of 2015, 64% of the overall US population and 82% of persons aged 18-49 years owned a smartphone [18]. Additionally, 15% of the population now owns a mobile phone-connected wearable device, such as a Fitbit or smartwatch [16]. As such, it is not surprising that mobile phone apps, that focus on health, fitness, or medical care (i.e., health apps), have become highly popular. Nonetheless, many cited concerns exist about app fees, privacy, and waning interest over time as reasons for letting the health apps lie fallow. There is a lack of usability testing of the apps, which eventually leads to burden of data entry and app abandonment. While app features can harness users' social instincts to help keep them diligent about tracking their health, when things go wrong they actually have the opposite effect. Existing apps provide some good features that most users appreciate, or some that users hate, and other that they would appreciate if it existed or done differently. Users often get frustrated by the huge variation in what information is available about foods' caloric content and nutritional features. There exist many good, bad and ugly features health tracking apps pose, and this is the main point we investigate with this study. We analyzed users review of 20 health tracking apps on AppStore and Google Play. Based on user's sentiment in the review we tried to investigate why certain apps were perceived as good or bad.

We classified the review into three distinct categories using the sentiment treebank with recursive neural tensor network. We then performed an inductive approach using the Gensim1 implementation of Word2Vec to infer the meaning of a word by measuring the cosine similarity. The Gensim package was ported from the original Word2Vec implementation by Google and extended with additional functionalities. We list the features as perceived by users and cluster similar feature under a high-level feature category and then provide an example describing each feature. We will highlight the positive, negative and neutral features described by users and provide future recommendation to consider. In Table-1, we list all the 20 apps analyzed together with their ranking, functionalities and other relevant information.

| Health Tracking Applications | | | | |
|---|---|---|---|---|
| **Applicaiton** | **Ratings** | **Price** | **Category** | **Bottom Line** |
| *FatSecret*[2] | 4.5 | Free | Health & Fitness | Calorie Counter |
| *MyNetDiary*[3] | 4.5 | 60$/year | Health & Fitness | Calorie Counter and Food Diary |
| *CalorieCounter*[4] | 4.5 | Free | Health & Fitness | Calorie Counter & Diet Tracker |
| *CRON-O-Meter Gold*[5] | 4.5 | 3,49e/month | Health & Fitness | Diet Tracker |
| *Eat This Much*[6] | 4.5 | Free | Health & Fitness | Meal Planner & Calorie Counter |
| *Food Planner Pal*[7] | 4.5 | Free | Food & Drinks | Food planner & diet tracker |
| *Google Fit*[8] | 4.5 | Free | Health & Fitness | Tracker |
| *HAPIcoach*[9] | 4.5 | Free | Health & Fitness | Nutrition Coaching |
| *Lifelog*[10] | 4.5 | 1,09e/month | Styles & Trends | Tracking your life |

---

| | | | | |
|---|---|---|---|---|
| *Lose It![11]* | 4 | Free | Health & Fitness | Calorie Counter and Weight Loss Tracker |
| *Mealime[12]* | 4.5 | Free | Food & Drinks | Healthy Meal Plans |
| *My Diet Diary[13]* | 4 | Free | Health & Fitness | Counter App |
| *MyDietCoach[14]* | 4.5 | Free | Health & Fitness | Weight Loss Booster, Calorie Counter |
| *MyFitnessPal[15]* | 4.5 | Free | Health & Fitness | Calorie Counter & Diet Tracker |
| *MyPlate[16]* | 4.5 | Free | Health & Fitness | Calorie Tracker |
| *Noom Coach[17]* | 4 | Free | Health & Fitness | Health and Weight |
| *RiseUp[18]* | 5 | Free | Health & Fitness | An Eating Disorder Monitoring and Management Tool for, Anorexia, Bulimia, Binge Eating, and EDNOS |
| *S Health[19]* | 4 | Free | Health & Fitness | Tracker |
| *Sweat with Kayla[20]* | 4 | Free | Health & Fitness | Bikini Body Fitness Workouts |
| *YAZIO[21]* | 4.5 | Free | Health & Fitness | Calorie Counter & Nutrition Tracker |

**Table-1: Health and fitness tracking applications.**

## 2. BACKGROUND

User reviews about an app can be crucial to understand aspects of app use and limitations [9]. Thus far, little work has been devoted to analyzing these reviews on app markets. Most studies have focused on app functionalities and usability and ignored user experience with these apps. For example, Frank et al. [8] collected a corpus of 188,389 Android apps from several Android app store to uncover patterns in the Android permission requests with a boolean matrix factorization. Studies in wearable fitness trackers have analyzed user sale posts on second sale e-commerce sites [6, 11]. A study by Clawson et al. analyzed advertisements of secondary sales of such technologies on Craigslist. The study investigated why users abandon personal health-tracking technologies. The study identified health motivations and rationales for abandonment and presented a set of design implications. Analyzing user attrition with mobile apps is a significant challenge, since there is little external pressure to use an app. Few studies exists about user's usage behaviour of health promotion apps. A web-based intervention found that adherence is lower outside randomized controlled trials and some observational studies have reported adherence rates as low as 1% [10]. Existing user ratings are summarized with simple histograms, and there aren't many analytical tools to provide insights about their review sentiment. However, a paper by Fu et al. [9] proposed WisCom, a system that can analyze user reviews and ratings on application markets. The tool can identify reasons why users like/dislike an app and provide valuable insights about users' major concerns and preferences of different types of apps. A similar study by Sangani et al., [17] analyzed user sentiments towards apps through their reviews and ratings. The study proposed a system that provides a many-to-many mapping from reviews to topics of interest, and a list of reviews per topic that are representative of user sentiment towards that topic.

Food journaling and health tracking is understood to be both important, yet difficult to sustain. Little is known about specific user challenges experienced during the app use. This leads to user burden and abandonment of the application. A study by Cordeiro et al. [7] identified key challenges in a qualitative study combining a survey of 141 current and lapsed food journal users with analysis of 5,526 posts in community forums for three mobile food journals. The study identified barriers to reliable food entry, negative nudges caused by current techniques, and challenges with social features. Mobile apps can make dietary self-monitoring easy with photography and potentially reach huge populations. However, the issue of sustained use of the app is still persisting. Helander et al. [13] conducted a retrospective analysis on the sample of 189,770 users who had downloaded the app and used it at least once. Adherence was defined based on frequency and duration of self-monitoring. People who had taken more than one picture were classified as "Users" and people with one or no pictures were classified as "Dropouts". Users who had taken at least 10 pictures and used the app for at least one week were classified as "Actives",

---

11 https://www.loseit.com
12 https://www.mealime.com
13 https://play.google.com/store/apps/details?id=org.medhelp.mydiet&hl=en
14 https://www.mydietcoachapp.com
15 https://www.myfitnesspal.com
16 https://www.choosemyplate.gov
17 https://www.noom.com
18 https://itunes.apple.com/us/app/rise-up-recover-eating-disorder-monitoring-management/id509287014?mt=8
19 https://www.samsung.com/us/support/owners/app/samsung-health
20 https://www.kaylaitsines.com
21 https://www.yazio.com

Users with 2-9 pictures were classified as "Semi-actives", and Dropouts with one picture were classified as "Non-actives" [13]. The study examined the association between adherence, registration time, dietary preferences, and peer feedback. Other studies [1, 12] focused on examining the content and features of apps for smoking cessation and association between feature use and quitting. The first study showed that quit plan, tracking, progress, and sharing features were mostly used, whereas the second study found that apps had low levels of adherence to key guidelines in the index. A similar study focused on characterizing the purpose and content of cancer focused apps and the evidence on their utility or effectiveness. A total of 295 apps from the smartphone app stores met the inclusion criteria. However, the study revealed lack of evidence on their utility, effectiveness, and safety [3]. The use of apps to assist with weight management is increasingly prevalent, but the quality of these apps is not well characterized. A study by Azar et al. [2] evaluated diet/nutrition and anthropometric tracking apps based on incorporation of features consistent with theories of behaviour change. The study conducted a comparative, descriptive assessment of top-rated free apps in the Health and Fitness category available in the iTunes App Store. All apps received low overall scores for inclusion of behavioral theory-based strategies. A similar study by West et al. [21] provided an overview of the developers' descriptions of health and fitness apps available on iTunes and appraises app's potential to influence behaviour change. The Health Education Curriculum Analysis Tool (HECAT) and the Precede-Proceed Model (PPM) were used as frameworks to guide the coding of 3336 paid apps. The result showed that apps exceeding US $0.99 were more likely to be scored as intending to promote health or prevent disease (92.55%, 1925/3336 vs 83.59%, 1411/3336), to be credible or trustworthy (91.11%, 1895/3336 vs 86.14%, 1454/3349), and more likely to be used personally or recommended to a health care client (72.93%, 1517/2644 vs 66.77%, 1127/2644). A study by Chen et al. [5] evaluated the quality of the most popular dietary weight-loss smartphone apps on the commercial market using comprehensive quality assessment criteria, and to quantify the behaviour change techniques (BCTs) incorporated. The accuracy of app energy intake calculations was further investigated by comparison with results from a 3-days weighed food record (WFR). Another study by West et al. [20] evaluated the extent to which diet apps' content was guided by health behaviour theory in their design and user interface. The study concluded that most apps were theory deficient and provided general information/assistance. Several health tracking apps exist, but it's unclear whether they adhere to evidence-based guidelines. Breland et al. [4] conducted app review on AppStore on diabetes apps. The study showed most diabetes apps do not conform to evidence-based recommendations, and future app reviews would benefit from testing app performance [4].

## 3. METHOD

We followed a screening process to select the apps with highest ratings by 2018 (=>*4-star rating*). The process included searching two apps from Google Play and the rest from App Store. We then extracted information about each app from the application market store on either platform. Next, we selected the most recent 100 user reviews for each app. This followed a filtering process using the Sentiment Treebank with Recursive Neural Tensor Network [19] and Word2Vec to code the reviews into good, bad and ugly.

To classify the reviews, we ran all the reviews through the Sentiment Treebank with Recursive Neural Tensor Network [19] to obtain and compare the sentiment tree returned with our analysis. This included classifying the review into very negative, negative, neutral, positive, and very positive. Where the negative is associated with bad review, the positive with good review, and the neutral is often associated with ugly review. Finally, we checked these sample of words in our Word2Vec model and selected the first 10 similar words returned per category and for all the apps (see Figure-1 for the apps screening pipeline process).

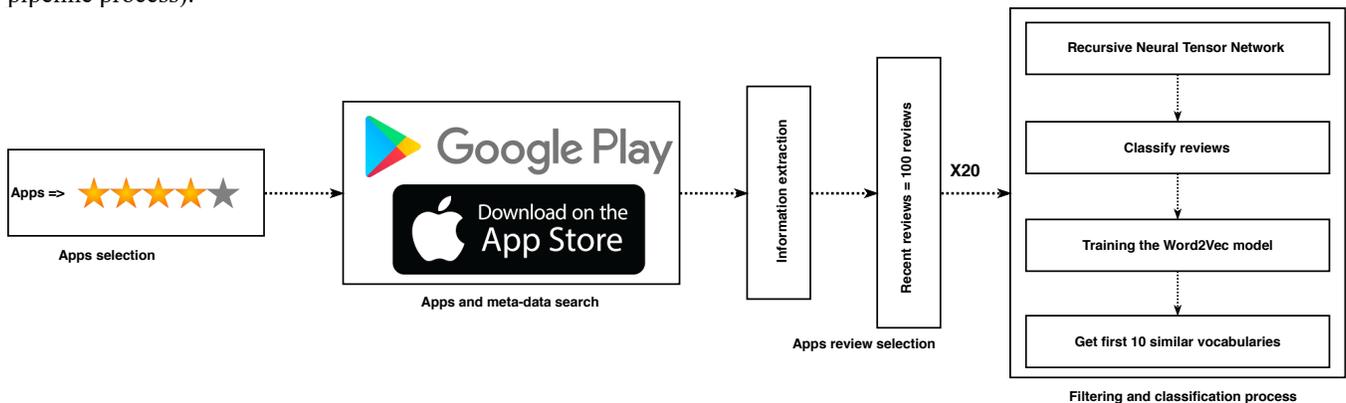

**Figure 1: The Application Selection and Processing Pipeline.**

## DATA COLLECTION AND TRAINING

We collected meta information and 100 users review of 20 diet, physical activity and other health tracking applications from both Google Play and Apple's AppStore in July 2018. To access information about these apps, we checked for their description on the app platforms. To build our dataset, we ran a review process of all the 20 apps on Google Play and AppStore and stored the first 100 reviews. The application meta-data included their name, rating, price, category and the description about their functionalities and purpose (all the meta-data analysis are summarized in Table-1). After accessing apps meta-data, we collected 100 reviews per app, with a total of 2000 user reviews for all the 20 apps. Each user review consists of a timestamp showing when the review was created, a user rating, and the user's comment. This dataset was then compressed and used to train the Word2Vec model. After we built the vocabulary we trained the Word2Vec model. This trains one simple neural network with a single hidden layer. The goal is to lean the weights of the hidden layer, which are the word vectors we're trying to learn. We used only reviews discussing either a good, a bad or an ugly feature in the app. The final list of reviews we considered in our analysis was 1022 reviews from all the three categories.

**PARAMETER SETTINGS**

**size:** The size of the dense vector that is to represent each token or word. We used a value of 150 dense vector representation.

**window:** The maximum distance between the target word and its neighboring word. The default value of 10 was used in our model.

**min_count:** The minimum frequency count of words. The model would ignore words that do not satisfy the min_count. We cleaned the data from infrequent words.

**workers:** This is the number of threads to use behind the scenes. We used 10 workers to train our model.

## 4. REVIEW ANALYSIS

We analyze the reviews obtained from both app platforms based on user experience with the app and the sentiment analysis of their review. The review analysis checked for positive, negative and neutral words into the review to first understand user motives from the review and then collect information about the specific feature they discuss. For each feature review, we collect and show the words relevant to that feature type. The review analysis consisted of three steps, we first ran the reviews into the Sentiment Treebank with Recursive Neural Tensor Network [19] to confirm our analysis and classification of the reviews. We then checked the review keywords for each of the categories with the Word2Vec model to obtain the first 10 keywords appeared in each review category. Finally, we analyzed the frequency of these keywords in the overall reviews. Words with unrecognized characters, capital letters, typos, slangs, or non-English were all pre-processed using the *gensim.utils.simple_preprocess ()*. This was necessary to tokenize and return a list of tokens (words).

**The Good**. The good features often return positive, or very positive results when analyzed with the Sentiment Treebank. We wanted to quantitatively and qualitatively determine if users are praising an app or complaining about it, or even being neutral or wished for some aspects in the app. The Sentiment Treebank with Recursive Neural Tensor Network was used to confirm positive reviews as good reviews and build a tree-like structure of the sentences. In Figure-2 we provide an example of positive sentiment treebank for a good user review. Based on the word tokens extracted from the good reviews, we built a vocabulary bag of words and checked the 10 most frequent words in the dataset on the Word2Vec model. These words were used to understand specific praising about features found in the apps, see Figure-3 for the vocabulary list and their frequency.

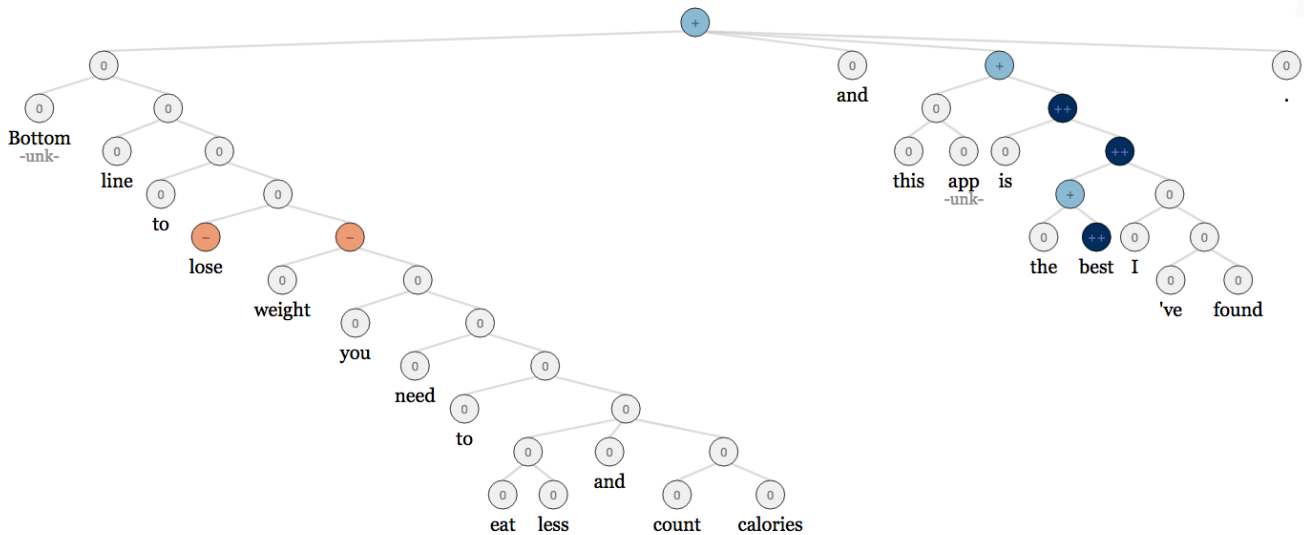

**Figure 2: A positive sentiment treebank example ©.**

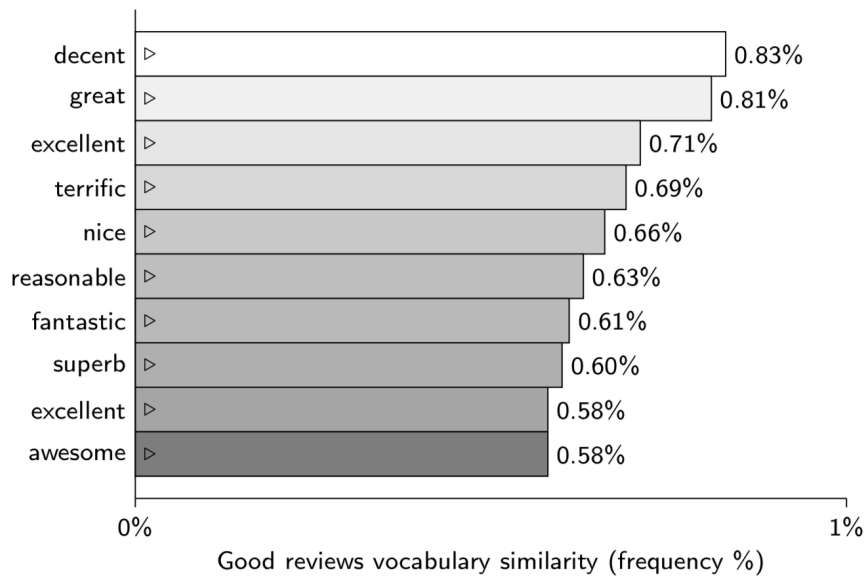

**Figure 3: Word2Vec – 10 most similar positive vocabularies.**

**The Bad.** The bad features where users discuss a certain technical, design or interaction issues about the app. The Sentiment Treebank with Recursive Neural Tensor Network was used to confirm negative reviews as bad reviews and build a tree-like structure of the sentences. In Figure-4 we provide an example of negative sentiment treebank for a bad user review. Based on the word tokens extracted from the bad reviews, we built a vocabulary bag of words by checking them on the Word2Vec model. These words were used to understand specific complaints about features found in the apps, see Figure-5 for the vocabulary list and their frequency.

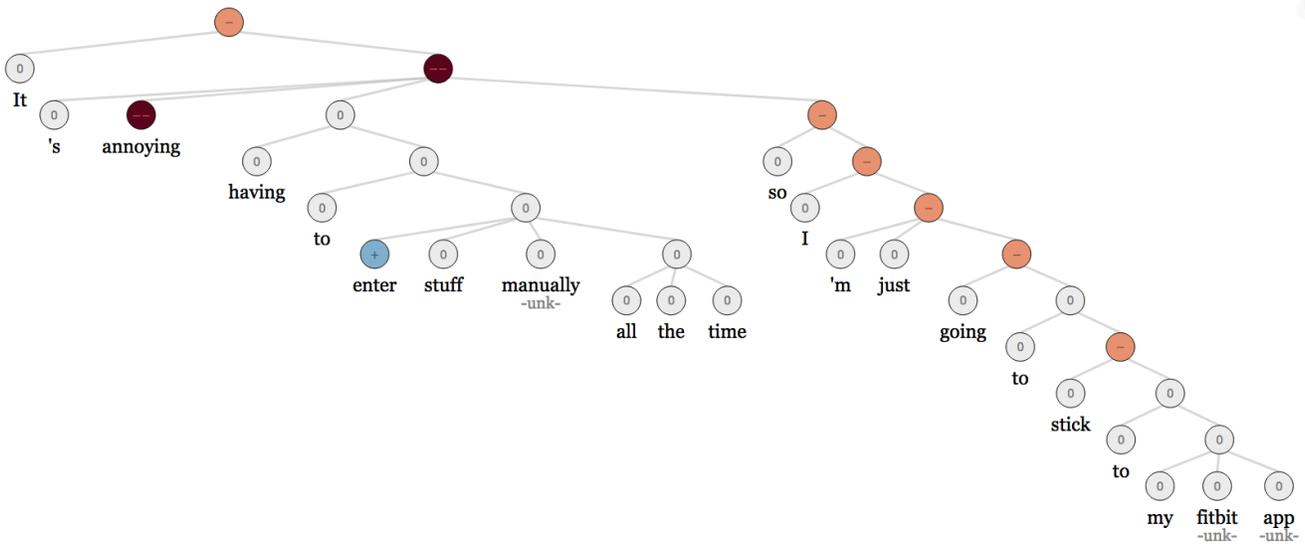

**Figure 4: A negative sentiment treebank example ©.**

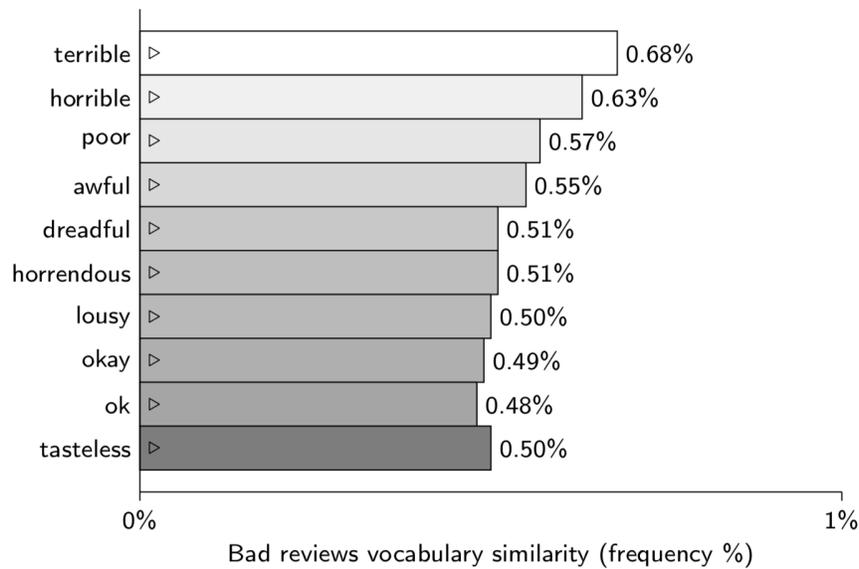

**Figure 5: Word2Vec – 10 most similar negative vocabularies.**

**The Ugly.** The ugly features where users discuss a certain technical, design or interaction requirement about the app. The discussion in this type of reviews is mostly neutral in terms of app perceiving, or the users often do not show their perceived positiveness or negativeness of the application. This includes reviews that wished to have a certain feature, design, interaction, or data presentation within the application. We followed the same analysis to extract this type of reviews from the data and based on specific feature discussed, we listed a column of feature types, the associated words, the frequency where these features appeared.

Sentiment Treebank with Recursive Neural Tensor Network was used to confirm neutral reviews as ugly reviews and build a tree-like structure of the sentences. In Figure-6 we provide an example of neutral sentiment treebank to illustrate an ugly review given by user. Based on the word tokens extracted from the bad reviews, we built a vocabulary bag of words by checking them on the Word2Vec model. These words were used to understand specific complaints about features found in the apps, see Figure-7 for the vocabulary list and their frequency.

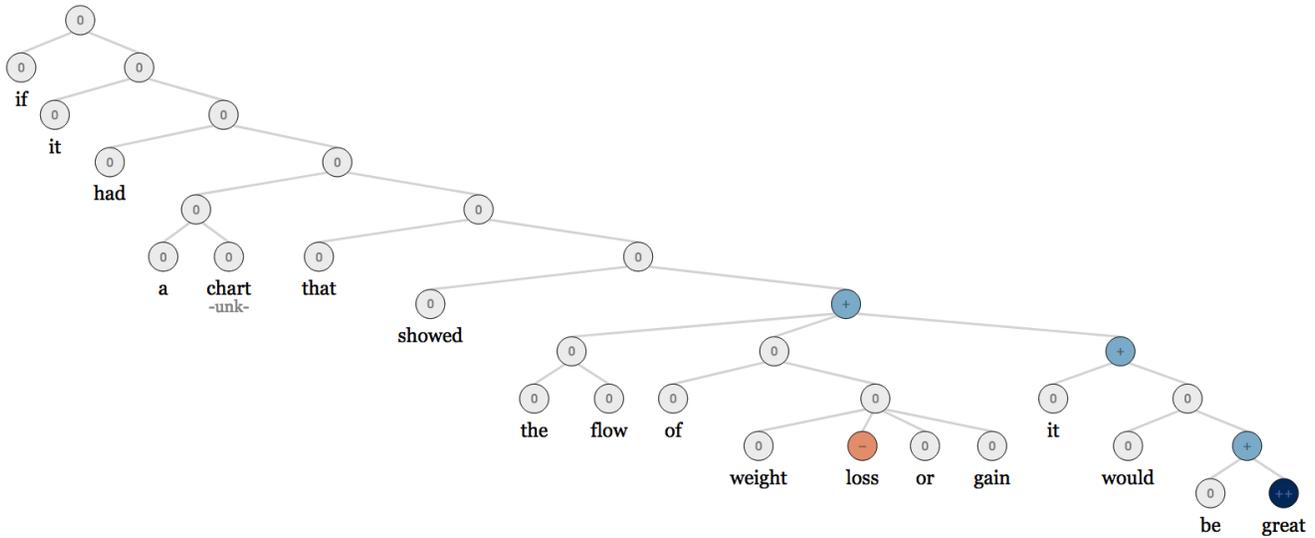

**Figure 6: A neutral sentiment treebank example ©.**

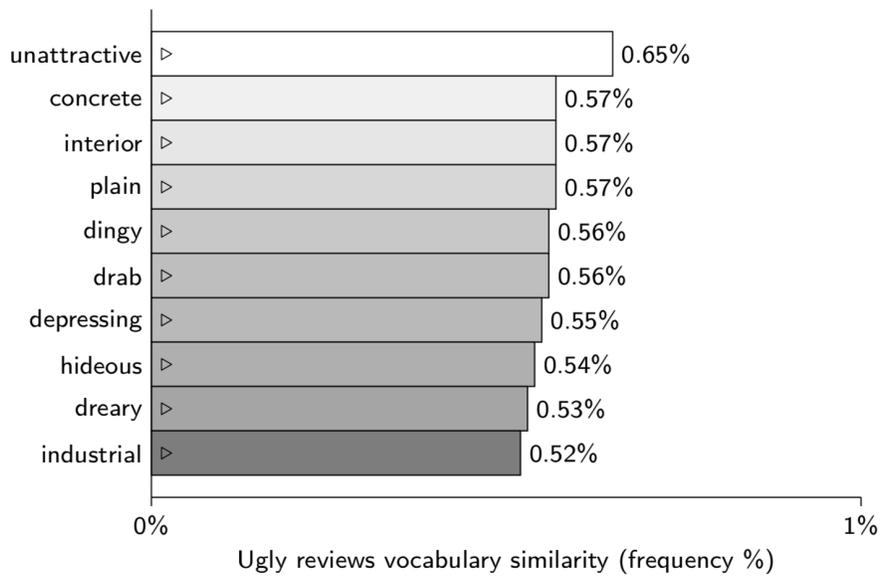

**Figure 7: Word2Vec – 10 most similar neutral vocabularies.**

## 5. RESULTS

We obtained reasonable results from the review data analysis, although we considered only the first 10 most frequent words provided by our Word2Vec model. The model returned similarity on the adjectives for each of the three sentiment categories. Our result confirmed all the words returned are used in the same context for a given query word obtained from the reviews. This similarity was measured by the cosine similarity function calculated inside our model. The main findings from the review analysis are discussed below.

### THE GOOD FEATURES

The positive features obtained from user's reviews were mostly about Flexibility, Simplicity, Performance, Customization, Efficiency, Entertaining, Social, Satisfaction, Helpful and Useful. From all the 20 apps analyzed, we obtained a total of 519 reviews praising a feature in an app. For a list of review examples for good features type, refer to Table-2 below.

Among the good reviews, most users discussed the simplicity aspect, with 87 reviews mentioning the simplicity, ease of use of an app. This was followed by review discussing user satisfaction with the app, with 79 reviews. Then the other most appreciated features were customization and flexibility, with 78 and 72 reviews, respectively. On the other hand, the social, entertaining, and performance aspects were among the least discussed good features of the apps, with 5, 7, and 7 reviews, respectively.

| The Good Features Type | |
|---|---|
| **Features** | **Sample review example** |
| *Flexibility* | Shows me how much I've lost over the years' time |
| *Simplicity* | It took me a short time to, learn to utilize this app |
| *Performance* | Very Improved The app has improved over the last year. It contains all of the features of the site and is far from lacking now. |
| *Customization* | Finally, a calorie app that lets me enter custom daily calorie goal! I am on HCG diet supervised by MD and is super low calorie daily. |
| *Efficiency* | I really like this app for tracking calories and workouts. |
| *Entertaining* | Amazing! Loved it! It makes the journey more fun :) besides, it is very helpful. |
| *Social* | I love this app even more! I love the options of support from other users as well! I am only 2 days in, but really like it so far. |
| *Satisfaction* | Best calorie counter & weight tracker! |
| *Helpful* | This app certainly helps, keep track of intake and output of calories. |
| *Useful* | Working well for me Just started using, so far very useful. |

Table 2: The Good Features Type Examples.

## THE BAD FEATURES

The bad feature type discussed by users were mostly about Bugs, Data inaccuracy, Customization, Price, Undesirable Techniques, GUI, Data limitations, Upgrade issues, UX issues, Undesirable Features, Hardware Limitations and Poor Customer Support. From all the apps reviewed, we obtained a total of 297 reviews complaining about a feature within an app. For a list of user reviews per feature, refer to Table-3 below. Among the bad reviews, most users discussed the issue of persisting bugs with the app, with 91 reviews complaining about continuous, unresolved bugs with the apps. This was followed by the issue with tracking and data accuracy which was mentioned by 34 reviews. Among the least complaint features were Hardware Limitations, Poor Customer Support, and Data limitations with 3, 11, and 12 reviews, respectively.

| The Bad Features Type | |
|---|---|
| **Features** | **Sample review example** |
| *Bugs* | Too many bugs not getting fixed Used to be my favorite up, Weeks have gone by without any pics from the developer. |
| *Data inaccuracy* | inaccuracy in steps counting |
| *Customization* | Doesn't seem to have ANY additional features over the regular app. Looks like gold features are on the website only. cant even create my own list! |
| *Price* | The app is 54 for three months. Add that to the books I already purchased from her, and it's well over what I would spend in a gym per year. |
| *Undesirable techniques* | This app is a con. Advertised as a free fitness community, it asks you to sign up for an account and once you've submitted your name and email address etc it tells you that it actually costs £15 |
| *GUI* | You know it's bad when you want to add "eggs and bacon" to your plan but can't without making a recipe for them. Seriously, not everything is a complicated recipe. |
| *Data limitations* | Food database still lacking. |
| *Upgrade issues* | Used to think this app was really good, but since the latest update the step counter doesn't work as well as it did before. |
| *UX issues* | The interface just has a cheap feel and the meal images never get formatted to fit well. |
| *Undesirable features* | I found out that items seen in the web browser wouldn't show up on the calendar view in the app, rendering the mobile app virtually useless. |
| *Hardware* | Play Services eat too much battery power. |
| *limitations* | |
| *Poor customer support* | I emailed the error that I am getting and have had no response. |

**Table 3: The Bad Features Type Examples.**

**THE UGLY FEATURES**

The ugly feature types discussed by users were mostly about Automation, Tracking, Extra Features, GUI, App name, Data limitations, Integration, Price, Undesirable Features, Bugs and Hardware. We obtained a total of 206 reviews being almost neutral about the app features from all the 20 apps analyzed. However, this type of reviews still expresses user's opinion about an app, although they always express their partial satisfaction with the app and wish for some changes. For a list of review examples per ugly features type, refer to Table-4 below. From all the ugly reviews, most users discussed the extra features they would like to have in an app, with 57 reviews. The other features followed were tracking and app bugs, with 27 and 25 reviews, respectively. The rest of the least discussed ugly features were automation, app name, and hardware issues, with 7, 2 and 2 reviews, respectively.

As an additional feature, we tracked the timeframe mentioned within the reviews to understand the approximate timeframe of use per application. Although the majority of reviews mentioned no timeframes of their use, however there were some reviews mentioned the timeframe of use, especially among the good and bad reviews as a sign of appreciation or frustration. For example, a user who wrote a positive review stated: *"I like the info-graphics. When I use this app, I can lose weight and keep on track. A few years and have seen it improve"*. Another user who wrote a negative review stated: *"I have been using the app for a few months now. The interface just has a cheap feel and the meal images never get formatted to fit well"*. The usage duration extracted from all the reviews lasted from a few weeks to a few years of use.

| The Ugly Features Type | |
|---|---|
| **Features** | **Sample review example** |
| Automation | 2 features I wish it had. 1. Automatically calculating net carbs. I would like to see my net carbs in my macros. The pie chart is "off" because it shows gross. |
| Tracking | When it works it is great, but it doesn't seem to pick up cycling and after 8 hours of wandering around a city it recorded only 44 minutes of walking. |
| Extra Features | but would like to be able to have multiple sleep records in a day. For instance, if you wake up in the middle of the night and stay awake for an hour or two |
| GUI | Navigating the app seems very difficult |
| App name | The name of the app however sounds a little negative though |
| Data limitations | Only thing is I've just ordered a Sony SmartBand as I can wear when I swim but don't see section in the software to record this |
| Integration | Needs better integration with S health for calorie burn and steps. |
| Price | I think the price is exaggerated |
| Undesirable features | It is very good to see how much I exercise but it doesn't know when I am in the car or running,It's more annoying |
| Bugs | I like the app and coach but it keeps on hanging (for android users) |
| Hardware | No integration, and sucks my battery... |

**Table 4: The Ugly Features Type Examples.**

**THE COMMON FEATURE**

From the review analyses, we found that some feature types appeared in more than a category, however, they carried a different sentiment meaning. For example, features such as customization, price, undesired features, or hardware limitation appeared in more than a category. Nonetheless, they carried different sentiment meaning or weight. For example, a GUI review on the bad feature category discusses about complexity of interacting with the app, as in: *"I found inserting and tracking the food is a tedious process"*, whereas the same category with ugly reviews might discuss a less serious GUI issue, as in: *"I'd love to see more advanced elements and better graphics on the app, so far its basic"*. In Figure-8 below we list the common features, their main categories and the intended sentimental meaning extracted in that category.

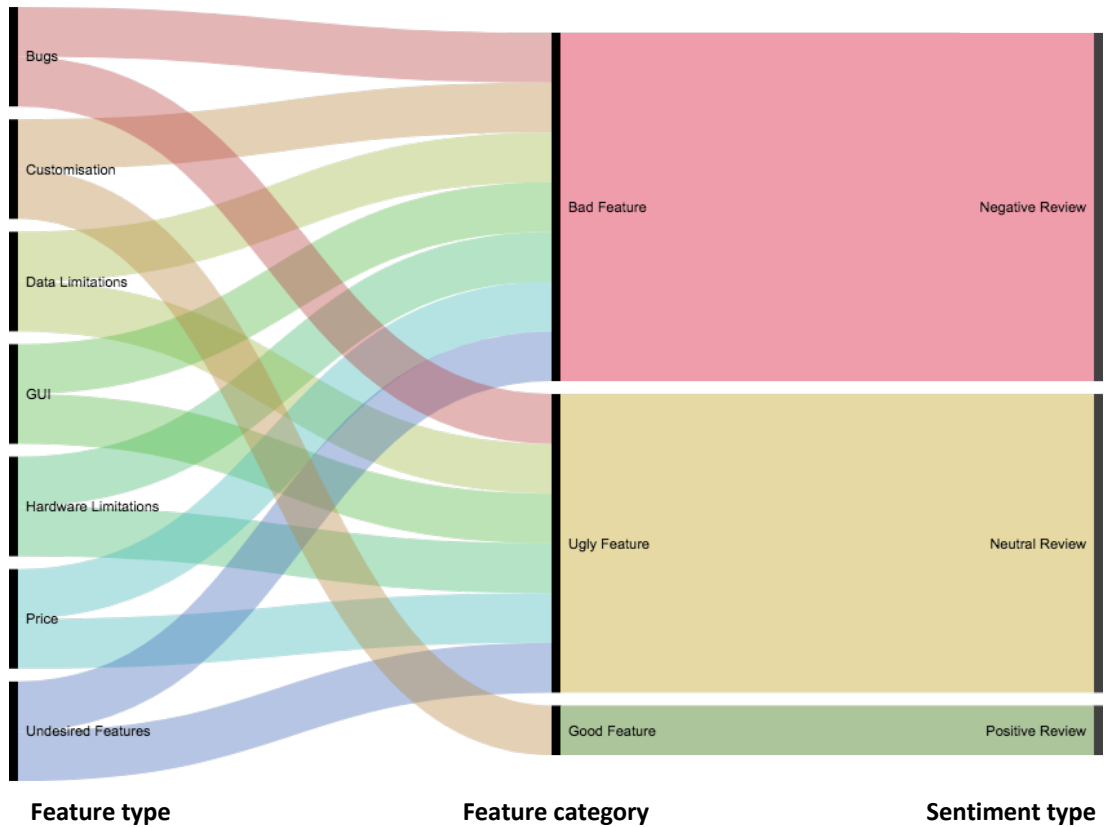

**Figure 8: Common features appeared in all three categories.**

## 6. DISCUSSION AND IMPLICATIONS

We explored user reviews about application features among the mHealth applications for diet, physical activities and other health aspects tracking. The results revealed certain feature patterns exist among all apps and some feature types are perceived more important than others. Among all the 20 apps and 100 reviews per app, some applications had more of a certain feature type than others, whereas there were some apps that had a combination of the feature types. The findings highlight the need for rigorous evaluation of the effectiveness of these apps and the effect of their features discussed in the review. However, it is possible that apps that have feature appealing to users might yet be ineffective for the health intervention.

Considering user reviews are important to detect strong and weak points in the application. Defining an approach to extract these reviews and specific features discussed helps to understand what experiences users have with the app. The good features are characterized by defining the features satisfying user needs, which in the long-term indicates apps success in triggering the users. The bad features are characterized by features either not targeting a specific user group or are developed with the wrong targeted group. However, bad features could also be due to poor design, bad implementation or even poor business model. The ugly features are characterized by being not completely satisfying or frustrating. This means that such features are usually weak or shaky and require more analysis and investigation. Our evaluation has shown that previous scientific literature provides practical analysis of user's interaction with the apps and analyzing their experience with these applications. Few studies performed a systematic analysis on user reviews rather than application reviews to understand reasons behind user appreciation or frustration with a certain application.

## 7. LIMITATIONS

This paper contributes to the existing research on health tracking and promotion applications by enabling designers and developers to efficiently process, manipulate and apply best practices extracted from user reviews. Promising research directions include dynamically analyzing user interaction with an application based on their reviews.

Results of this study should be interpreted in light of key limitations. First, the findings from user review analysis should be interpreted with caution because, having users discussing a feature deems interesting, but it doesn't always reveal a good or bad user experience with the application. Second, associations between feature discussed and user experience should be correlational and shouldn't contradict. Additional studies are needed to experimentally manipulate the availability of app features that responds well to new user needs. Third, the structure and appearance of an app, or of any other technology-delivered intervention, can affect user behaviour. Thus, results of this study should be interpreted in light of the possibility that feature usage may have also been influenced by factors other than interest in specific content (e.g., app design and aesthetics).

## 8. CONCLUSION

There is a big disruption with personal health tracking technologies, as they are rapidly adopted into mainstream culture and have sparked an explosion of interest in tracking various health aspects. However, these technologies suffer from being abandoned in the long-term. In this work, we analyzed 20 health tracking applications that are considered among the highly rated apps in both AppStore and Google Play, we then collected the most recent 100 user reviews per application. The analysis focused on classifying and clustering the reviews into good, bad and ugly reviews. We were able to detect different usage patterns among users based on their review. We performed a sentiment analysis on the reviews after we obtained the results using the Sentiment Treebank with Recursive Neural Tensor Network. The result confirmed the positive, negative and neutral aspect of the reviews. We were able to identify the major reasons why users dislike an app and learn how users' complaints was related to a specific feature. We coded these features into a high-level feature category. The findings revealed patterns of what feature category was perceived as positive, negative or neutral by the users. Future work will apply in-depth analysis to investigate different review patterns triggered by user experience with the application and to what extend this could be used as a guideline for future development.

## ACKNOWLEDGMENT
We acknowledge that the graphs in Figures-2,4,6 are produced by the Sentiment Treebank with Recursive Neural Tensor Network tool.